\documentclass[aps,twocolumn,prd,amsmath,nofootinbib,preprintnumbers,superscriptaddress]{revtex4-1}
\usepackage{epsfig,slashed,nicefrac}
\usepackage[utf8]{inputenc}
\usepackage{color}
\usepackage{float} 
\usepackage{subfig} 
\usepackage{eqnarray,amsmath}
\usepackage{multirow}

\def\be{\begin{equation*}}
\def\ee{\end{equation*}}
\def\bsp#1\esp{\begin{split}#1\end{split}} 
\def\bpm{\begin{pmatrix}}
\def\epm{\end{pmatrix}}

\begin{document}


\title{Triple Prompt $J/\psi$ Hadroproduction as a Hard Probe of Multiple-parton scatterings}

\author{Hua-Sheng Shao}
\affiliation{Laboratoire de Physique Th\'eorique et Hautes Energies (LPTHE), UMR 7589, Sorbonne Universit\'e et CNRS, 4 place Jussieu, 75252 Paris Cedex 05, France}

\author{Yu-Jie Zhang}
\affiliation{School of Physics and Nuclear Energy Engineering, Beihang University, Beijing 100083, China}
\affiliation{Center for High Energy Physics, Peking University, Beijing 100871, China}

\date{\today}

\begin{abstract}
We propose that the process of triple prompt $J/\psi$ hadroproduction is a very clean hard probe of multiple-parton scatterings at high-energy hadron colliders, especially the least known triple-parton scattering. A first complete study is carried out by considering single-, double-, and triple-parton scatterings coherently. Our calculation shows that it is a golden channel to probe double- and triple-parton scatterings, as the single-parton scattering is strongly suppressed. The predictions of the (differential) cross sections in proton-proton collisions at the LHC and the future higher-energy hadron colliders are given. Our study shows that its measurement is already feasible with the existing data collected during the period of the LHC run 2. A method is proposed to extract the triple-parton scattering contribution, and therefore it paves a way to study the possible triple-parton correlations in a proton.
\end{abstract}

\maketitle

\textit{Introduction} -- Multiparton scattering (MPI) physics at the high-energy hadron colliders, like the Large Hadron Collider (LHC) at CERN and future hadron colliders~\cite{Benedikt:2018ofy},  is becoming increasingly important to study the new phenomenon~\cite{Doncheski:1993dj,Dunnen:2014eta} in the standard model and to search for beyond the standard model signatures~\cite{Grifols:1987iq,Robinett:1991us,Clarke:2013aya,Acosta:2003mu,Aaltonen:2014rda,Aad:2014rua,Aad:2014kba,Lansberg:2017dzg} with the fast increase of the parton-parton luminosity. As opposed to the leading MPI double-parton hard scattering (DPS), the measurements of the next-to-leading MPI triple-parton scattering (TPS) at the LHC are absent due to their more complicated final states and much fewer yields. Such rare processes, however, are possible to study with enough statistics at the high-luminosity phase of the LHC (HL-LHC). Similar to the DPS case, the general factorization ansatz of MPI exists~\cite{dEnterria:2017yhd} and perturbative QCD (PQCD) calculations are possible~\cite{Blok:2011bu,Diehl:2011yj,Diehl:2011tt,Gaunt:2011xd,Manohar:2012pe,Gaunt:2012dd,Blok:2013bpa,Diehl:2014vaa,Diehl:2015bca,Rinaldi:2016jvu,Buffing:2017mqm,Diehl:2017kgu,Vladimirov:2017ksc,Diehl:2018wfy,Gaunt:2018eix} given the full (yet known) knowledge of multiple-dimensional tomography of the proton. In practice, the phenomenological studies of MPI are either strongly model dependent or assuming no correlation between the multiple-parton scatterings. We will use the latter approach here as a testable ground to study the multiple-parton correlations from MPI. The current DPS studies at the LHC and Tevatron suggest that the zero correlation assumption is a rather good approximation. With this assumption, a generic N-parton scattering (NPS) cross-section becomes~\cite{dEnterria:2017yhd}
\begin{eqnarray}
\sigma^{\rm NPS}_{f_1\cdots f_N}&=&\frac{m}{N!}\frac{\prod_{i=1}^{N}{\sigma^{\rm SPS}_{f_i}}}{\left(\sigma_{\rm eff,N}\right)^{N-1}},\label{eq:NPSxs}
\end{eqnarray}
where the combinatorial factor $\frac{m}{N!}$ takes into account the indistinguishable final state symmetry and and $\sigma^{\rm SPS}_{f_i}$ is the single-parton scattering (SPS) cross section of producing final state $f_i$. The effective cross section $\sigma_{\rm eff,N}$ encodes all possible unknown parton transverse profiles in the protons, and should be determined by experiments. The DPS and TPS cases correspond to $N=2$ and $N=3$ in the above formula Eq.(\ref{eq:NPSxs}). From the pure geometrical consideration, Ref.~\cite{dEnterria:2016ids} derives $\sigma_{\rm eff, 3}=\left(0.82\pm0.11\right)\times \sigma_{\rm eff,2}$ after a global survey of various parton transverse profiles.

Heavy quarkonia, bound states of heavy-flavored quarks, provide crucial insights of gluon-gluon and gluon-quark correlations in the proton by studying their associated production processes~\cite{Aaij:2011yc,Lansberg:2013qka,Lansberg:2014swa,Abazov:2014qba,Khachatryan:2014iia,Lansberg:2015lva,Aaboud:2016fzt,Aaij:2016bqq,Abazov:2015fbl,Shao:2016wor,Lansberg:2016rcx,Khachatryan:2016ydm,Lansberg:2016muq,Lansberg:2017chq,Aad:2014rua,Aad:2014kba,Aaij:2012dz,Aaij:2015wpa} in a wide kinematic range. The values of $\sigma_{\rm eff,2}$ for DPS extracted from the quarkonium data are in general smaller than 10 mb as opposed to 15 mb from other final states at higher scales, like the weak gauge boson processes~\cite{Aad:2013bjm,Chatrchyan:2013xxa,Sirunyan:2017hlu,Cao:2017bcb}. However, we should bear in mind that it is still far from being conclusive in view of the remaining large uncertainties.

On the other hand, TPS theoretical studies in literature are limited to open heavy-flavor productions~\cite{dEnterria:2016ids,Maciula:2017meb,Maciula:2017wpe} so far. Their complete study by including SPS and DPS is not available. In this Letter, we consider triple-$J/\psi$ hadroproduction as a TPS-case study and perform a first complete study by including SPS, DPS and TPS simultaneously. 

\medskip

\begin{figure}[hbt!]
\centering
\subfloat[SPS]{\includegraphics[width=.33\columnwidth,draft=false]{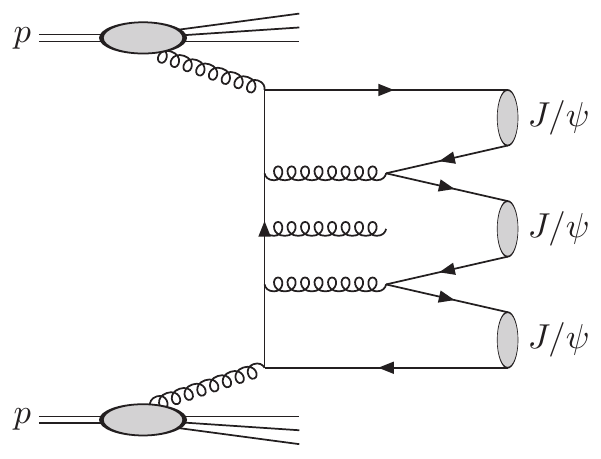}\label{diagram-SPS}}
\subfloat[DPS]{\includegraphics[width=.33\columnwidth,draft=false]{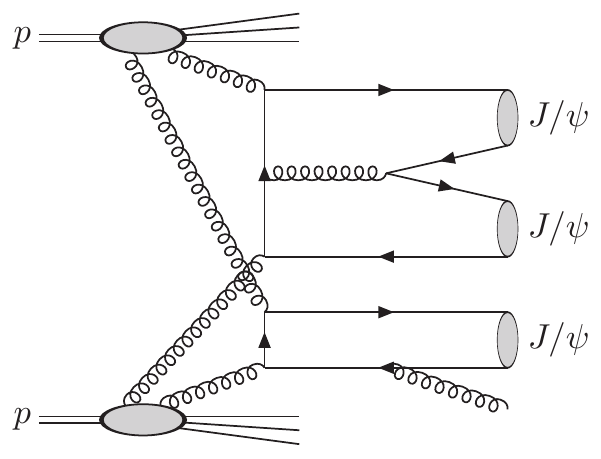}\label{diagram-DPS}}
\subfloat[TPS]{\includegraphics[width=.33\columnwidth,draft=false]{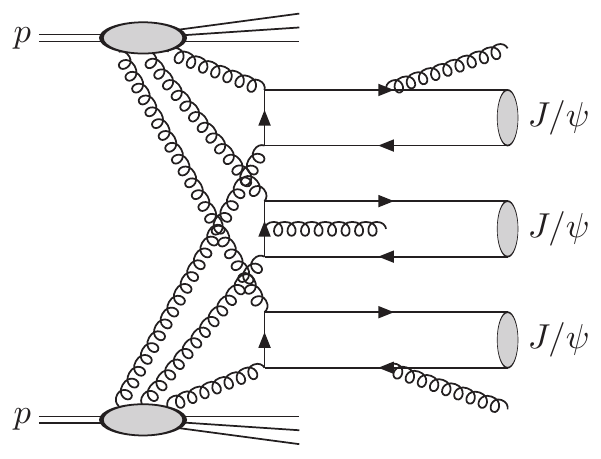}\label{diagram-TPS}}
\caption{Typical Feynman diagrams for triple $J/\psi$ hadroproduction via (a) SPS, (b) DPS and (c) TPS processes.}
\label{diagrams}
\end{figure}

\textit{Theoretical framework} -- In triple $J/\psi$ hadroproduction, there are three scattering processes (SPS, DPS and TPS) entering into the calculations of the (differential) cross sections, where we have shown one typical Feynman diagram for each mode in Fig.~\ref{diagrams}. Under the zero correlation assumption (\ref{eq:NPSxs}), we will use the following concrete formula:
\begin{eqnarray}
&&\sigma^{\rm DPS}(pp\rightarrow J/\psi J/\psi J/\psi+X)\nonumber\\
&=&\frac{\sigma^{\rm SPS}(pp\rightarrow J/\psi J/\psi+X)\sigma^{\rm SPS}(pp\rightarrow J/\psi+X)}{\sigma_{\rm eff,2}},\nonumber\\
&&\sigma^{\rm TPS}(pp\rightarrow J/\psi J/\psi J/\psi+X)\nonumber\\
&=&\frac{1}{6}\frac{\left[\sigma^{\rm SPS}(pp\rightarrow J/\psi+X)\right]^3}{\left(\sigma_{\rm eff,3}\right)^2}
\end{eqnarray}
to calculate DPS and TPS cross sections. In total, there are three different SPS cross sections, i.e., those of one, two and three $J/\psi$ production, to be computed. A similar hybrid approach proposed in Refs.~\cite{Lansberg:2014swa,Lansberg:2015lva,Shao:2016wor} will be adopted here. The matrix elements for double and triple prompt $J/\psi$ SPS productions are based on PQCD calculations in the non-relativistic QCD (NRQCD) factorization framework~\cite{Bodwin:1994jh}, while the single $J/\psi$ hadroproduction is estimated by the data-driven approach. 

The SPS cross sections for single and double $J/\psi$ production have been extensively studied in literature. One encounters the difficulties in understanding single $J/\psi$ production in NRQCD, especially for the subleading color-octet channels. Given the availability of its precision measurements at the LHC covering a wide kinematic regime, we will use the data-driven approach to fit the matrix element of the single $J/\psi$ production with the precise experimental data~\cite{Lansberg:2014swa}.

On the other hand, we will use PQCD calculations to determine the SPS yields of two and three $J/\psi$ production. For a multiple quarkonium SPS production process, its cross-section can be written as
\begin{eqnarray}
&&\sigma^{\rm SPS}(pp \rightarrow \mathcal{Q}_1\cdots \mathcal{Q}_m+X)=\sum_{n_1,\cdots,n_m}\nonumber\\
&&\left[\hat{\sigma}^{\rm SPS}(pp\rightarrow Q\bar{Q}[n_1] \cdots Q\bar{Q}[n_m]+X)\prod_{i=1}^{m}{\langle \mathcal{O}^{\mathcal{Q}_i}(n_i)\rangle}\right],
\end{eqnarray}
where the long-distance matrix elements (LDMEs) $\langle \mathcal{O}^{\mathcal{Q}_i}(n_i)\rangle$ follow the power counting of NRQCD velocity scaling rule. The leading Fock state $Q\bar{Q}[n]$ for S-wave quarkonium, with the assumption of the same order of magnitude in the short-distance coefficients (SDCs) $\hat{\sigma}$, shares the same quantum number $J^{\rm PC}$ and color representation as the quarkonium. Depending on the kinematic region, the working assumption on the size of the SDCs may not always hold. A notorious example is the boosted single inclusive $J/\psi$ production, which receives giant K factors from QCD radiative corrections and is dominated by the subleading Fock states (see Ref.~\cite{Shao:2018adj} for a recent discussion). Therefore, one should always bear in mind to carefully check the working assumption case by case. It certainly complicates the corresponding quarkonium phenomenology studies.


The PQCD calculation of double $J/\psi$ at leading-order (LO) in $v^2$ (where $v$ is the relative velocity between two heavy quarks in the rest frame of the quarkonium) and next-to-leading order (NLO) in $\alpha_s$ shows a fairly good agreement with the data when its DPS is small~\cite{Aaboud:2016fzt} and/or after subtracting the estimated DPS~\cite{Aaij:2016bqq,Lansberg:2014swa}. In the present Letter, we will use partial NLO result of the double $J/\psi$ SPS part by including infrared-safe real emission diagrams only. It shows a reasonable agreement with the complete NLO calculation~\cite{Sun:2014gca}.


Besides, we perform a first calculation of the SPS cross-section for triple $J/\psi$ production here. As opposed to double $J/\psi$ production, the LO SDC in $\alpha_s$ at leading $v^2$ for single and triple $J/\psi$ production must accompany a hard gluon in the final states within the PQCD framework. In other words, it is $\mathcal{O}(\frac{\alpha_s}{v^4})$ compared to their subleading Fock state channels. The expected hierarchy of different Fock states should be respected as long as $J/\psi$ is not boosted, because $\frac{\alpha_s}{v^4}\approx \frac{1}{v^2}\gg 1$ given $\alpha_s\approx v^2\approx 0.2$. It is indeed the case in the single $J/\psi$ production. At low $P_T$ or in the $P_T$-integrated cross section, the leading Fock state $^3S_1^{[1]}$~\footnote{We have used the spectroscopy notation $^{2s+1}L_J^{[c]}$ here, where $s$ is the spin, $L$ is the orbital angular momentum, $J$ is the total angular momentum, and $c$ is the color representation of the heavy quark pair.} contribution can describe the experimental measurements of $J/\psi$ pretty well~\cite{Feng:2015cba}. We expect the situation in SPS triple $J/\psi$ production is analogous to the single $J/\psi$ case. Since triple $J/\psi$ production is a very rare process and we are interested in its discovery potential at the LHC and the future colliders, we will not push $J/\psi$ to the phase-space corners. It is expected that the leading $v^2$ partonic channel $gg\rightarrow c\bar{c}[^3S_1^{[1]}]+c\bar{c}[^3S_1^{[1]}]+c\bar{c}[^3S_1^{[1]}]+g$ works well as long as $P_T$ of $J/\psi$ is not large. However, even at LO, the process is already challenging enough on both sides of the scattering amplitude computations and the phase-space integrations. There are more than $2\cdot 10^{4}$ Feynman diagrams to be tackled. The computation is achieved here for the first time with the help of {\sc\small HELAC-Onia}~\cite{Shao:2012iz,Shao:2015vga}, due to the virtue of the recursion relations.

\medskip

\textit{Results} -- The numerical calculations for SPS, DPS, and TPS triple $J/\psi$ production are performed in the {\sc\small HELAC-Onia} framework. In PQCD parts for double and triple $J/\psi$ yields, we take the charm mass being $1.5$ GeV, and the central scale $\mu_0=\frac{H_T}{2}$, where $H_T$ is the sum of the transverse masses of the final states. We will also independently vary the renormalization scale $\mu_R$ and the factorization scale $\mu_F$ by a factor of 2, i.e., $\mu_{R/F}=\xi_{R/F}\mu_0$ with $\xi_{R/F}=0.5,1$, and $2$. It is conventionally used to estimate the missing higher order in $\alpha_s$, which is the dominant theoretical uncertainty. We choose the proton parton-distribution function as CT14NLO~\cite{Dulat:2015mca}. In double and triple $J/\psi$ SPS cross sections, we have also included the feed-down contribution from the excited state $\psi(2S)$. The corresponding LDMEs are estimated in a potential model via $\langle \mathcal{O}^{\mathcal{Q}}(^3S_1^{[1]})\rangle=\frac{9}{2\pi}|R^{\mathcal{Q}}(0)|^2$, where the squared wave functions at the origin are $|R^{J/\psi}(0)|^2=0.81$ GeV$^3$ and $|R^{\psi(2S)}(0)|^2=0.529$ GeV$^3$~\cite{Eichten:1995ch}. For the single prompt $J/\psi$ production cross section, we use the same ansatz of the averaged amplitude squared [Eq.(1) in Ref.~\cite{Lansberg:2016deg} ] and fit to the LHCb data measured at $\sqrt{s}=7$ and $8$ TeV~\cite{Aaij:2011jh,Aaij:2013yaa}. The final fitted parameters in the ansatz are listed in the top row of Table 1 in Ref.~\cite{Lansberg:2016deg}.

The inclusive total cross sections, as well as those in the LHCb forward rapidity acceptance $2.0<y_{J/\psi}<4.5$ and the ATLAS/CMS central rapidity acceptance $|y_{J/\psi}|<2.4$, are presented in Table~\ref{tab:xsprompt}. We have multiplied the branching ratio of $J/\psi$ into muon pairs in the cross-sections. Four different center-of-mass energies $\sqrt{s}=13, 27, 75$, and $100$ TeV are quoted to represent the LHC and the proposed future hadron colliders~\cite{Benedikt:2018ofy}. We have quoted two theoretical uncertainties in each SPS cross-section. The first one is the renormalization and factorization scale uncertainty, while the second one is the error from the Monte Carlo integration. In the DPS cross sections, we only show the uncertainty from the scale variations, because their Monte Carlo errors are negligible. We do not show any theoretical error for TPS, as the matrix element of the single $J/\psi$ is determined by the very precise experimental data. In general, the SPS contributions are several orders of magnitude smaller than DPS and TPS cross sections, as long as the unknown effective cross sections $\sigma_{\rm eff,2}$ and $\sigma_{\rm eff,3}$ are not significantly larger than the reference value of $10$ mb. Such a conclusion holds regardless of the center-of-mass energy $\sqrt{s}$ and the rapidity cuts on $J/\psi$. 


A few comments on the integrated luminosities at the LHC and future hadron colliders are in order before we move to estimate the expected number of events. ATLAS and CMS experiments have collected around $150$ fb$^{-1}$ during the period of LHC run 2 at $\sqrt{s}=13$ TeV, and the corresponding number for the LHCb experiment is $6$ fb$^{-1}$. There will be two phases for HL-LHC runs~\cite{Cerri:2018ypt}. (Strictly speaking, the nominal center-of-mass energy at HL-LHC is 14 TeV instead of 13 TeV. Since the two energies are very close, we do not expect any significant difference for the cross-sections.) During phase 1, LHCb aims to deliver $23$ fb$^{-1}$, and ATLAS and CMS aim to deliver 300 fb$^{-1}$. The integrated luminosity of LHCb (ATLAS and CMS) will increase to 300 fb$^{-1}$ (3 ab$^{-1}$) at phase 2. The nominal integrated luminosities for the future hadron colliders ($27$ TeV high-energy LHC~\cite{Cerri:2018ypt}, $75$ TeV super proton-proton collider~\cite{CEPC-SPPCStudyGroup:2015csa}, and $100$ TeV future circular collider~\cite{Mangano:2017tke}) are in the range of $10-20$ ab$^{-1}$.

\begin{table*}
\begin{normalsize}
\hspace{0cm}
\begin{tabular}{c|c|c|c|c}
\hline\rule{0pt}{3ex}
 & & inclusive & $2.0<y_{J/\psi}<4.5$ & $|y_{J/\psi}|<2.4$\\[1mm] 
\hline\hline\rule{0pt}{3ex}
\multirow{3}{*}{13 TeV}  & SPS & $0.41^{+2.4}_{-0.34}\pm0.0083$ &$(1.8^{+11}_{-1.5}\pm0.18)\times 10^{-2}$
  & $(8.7^{+56}_{-7.5}\pm0.098)\times 10^{-2}$\\
\cline{2-5}
\rule{0pt}{3ex}
& DPS  & $(190^{+501}_{-140})\times \frac{10~{\rm mb}}{\sigma_{\rm eff,2}}$ & $(7.0^{+18}_{-5.1})\times \frac{10~{\rm mb}}{\sigma_{\rm eff,2}}$
  & $(50^{+140}_{-37})\times \frac{10~{\rm mb}}{\sigma_{\rm eff,2}}$\\
\cline{2-5}
\rule{0pt}{3ex}
& TPS  & $130\times \left(\frac{10~{\rm mb}}{\sigma_{\rm eff,3}}\right)^2$ & $1.3\times \left(\frac{10~{\rm mb}}{\sigma_{\rm eff,3}}\right)^2$
  & $18\times \left(\frac{10~{\rm mb}}{\sigma_{\rm eff,3}}\right)^2$\\
\hline\rule{0pt}{3ex}
\multirow{3}{*}{27 TeV}  & SPS & $0.46^{+2.9}_{-0.39}\pm0.022$ &$(3.2^{+22}_{-2.8}\pm0.21)\times 10^{-2}$
  & $(5.8^{+39}_{-5.1}\pm0.29)\times 10^{-2}$\\
\cline{2-5}
\rule{0pt}{3ex}
& DPS  & $(560^{+2900}_{-480})\times \frac{10~{\rm mb}}{\sigma_{\rm eff,2}}$ & $(19^{+97}_{-16})\times \frac{10~{\rm mb}}{\sigma_{\rm eff,2}}$
  & $(120^{+630}_{-100})\times \frac{10~{\rm mb}}{\sigma_{\rm eff,2}}$\\
\cline{2-5}
\rule{0pt}{3ex}
& TPS  & $570\times \left(\frac{10~{\rm mb}}{\sigma_{\rm eff,3}}\right)^2$ & $5.0\times \left(\frac{10~{\rm mb}}{\sigma_{\rm eff,3}}\right)^2$
  & $57\times \left(\frac{10~{\rm mb}}{\sigma_{\rm eff,3}}\right)^2$\\
\hline\rule{0pt}{3ex}
\multirow{3}{*}{75 TeV}  & SPS & $0.59^{+4.4}_{-0.52}\pm 0.016$ &$(3.0^{+25}_{-2.7}\pm 0.23)\times 10^{-2}$
  & $(7.2^{+63}_{-6.5}\pm 0.38)\times 10^{-2}$\\
\cline{2-5}
\rule{0pt}{3ex}
& DPS  & $(1900^{+11000}_{-1600})\times \frac{10~{\rm mb}}{\sigma_{\rm eff,2}}$ & $(57^{+340}_{-50})\times \frac{10~{\rm mb}}{\sigma_{\rm eff,2}}$
  & $(310^{+2000}_{-270})\times \frac{10~{\rm mb}}{\sigma_{\rm eff,2}}$\\
\cline{2-5}
\rule{0pt}{3ex}
& TPS  & $3900\times \left(\frac{10~{\rm mb}}{\sigma_{\rm eff,3}}\right)^2$ & $27\times \left(\frac{10~{\rm mb}}{\sigma_{\rm eff,3}}\right)^2$
  & $260\times \left(\frac{10~{\rm mb}}{\sigma_{\rm eff,3}}\right)^2$\\
\hline\rule{0pt}{3ex}
\multirow{3}{*}{100 TeV}  & SPS & $1.1^{+8.4}_{-1.0}\pm0.044$ &$(4.5^{+33}_{-4.0}\pm0.72)\times 10^{-2}$
  & $(36^{+290}_{-32}\pm1.8)\times 10^{-2}$\\
\cline{2-5}
\rule{0pt}{3ex}
& DPS  & $(3400^{+19000}_{-2900})\times \frac{10~{\rm mb}}{\sigma_{\rm eff,2}}$ & $(100^{+550}_{-86})\times \frac{10~{\rm mb}}{\sigma_{\rm eff,2}}$
  & $(490^{+3000}_{-430})\times \frac{10~{\rm mb}}{\sigma_{\rm eff,2}}$\\
\cline{2-5}
\rule{0pt}{3ex}
& TPS  & $6500\times \left(\frac{10~{\rm mb}}{\sigma_{\rm eff,3}}\right)^2$ & $45\times \left(\frac{10~{\rm mb}}{\sigma_{\rm eff,3}}\right)^2$
  & $380\times \left(\frac{10~{\rm mb}}{\sigma_{\rm eff,3}}\right)^2$\\
\hline
\end{tabular}
\caption{Cross sections $\sigma(pp\rightarrow 3J/\psi)\times{\rm Br}^3(J/\psi\rightarrow \mu^+\mu^-)$ (in femtobarn) at $\sqrt{s}=13,27,75,100$ TeV proton-proton colliders, where we have also included feed-down contributions from higher-excited quarkonia decay.}
\label{tab:xsprompt} 
\end{normalsize}
\end{table*}

\begin{figure}[hbt!]
\centering
\includegraphics[width=.95\columnwidth,draft=false]{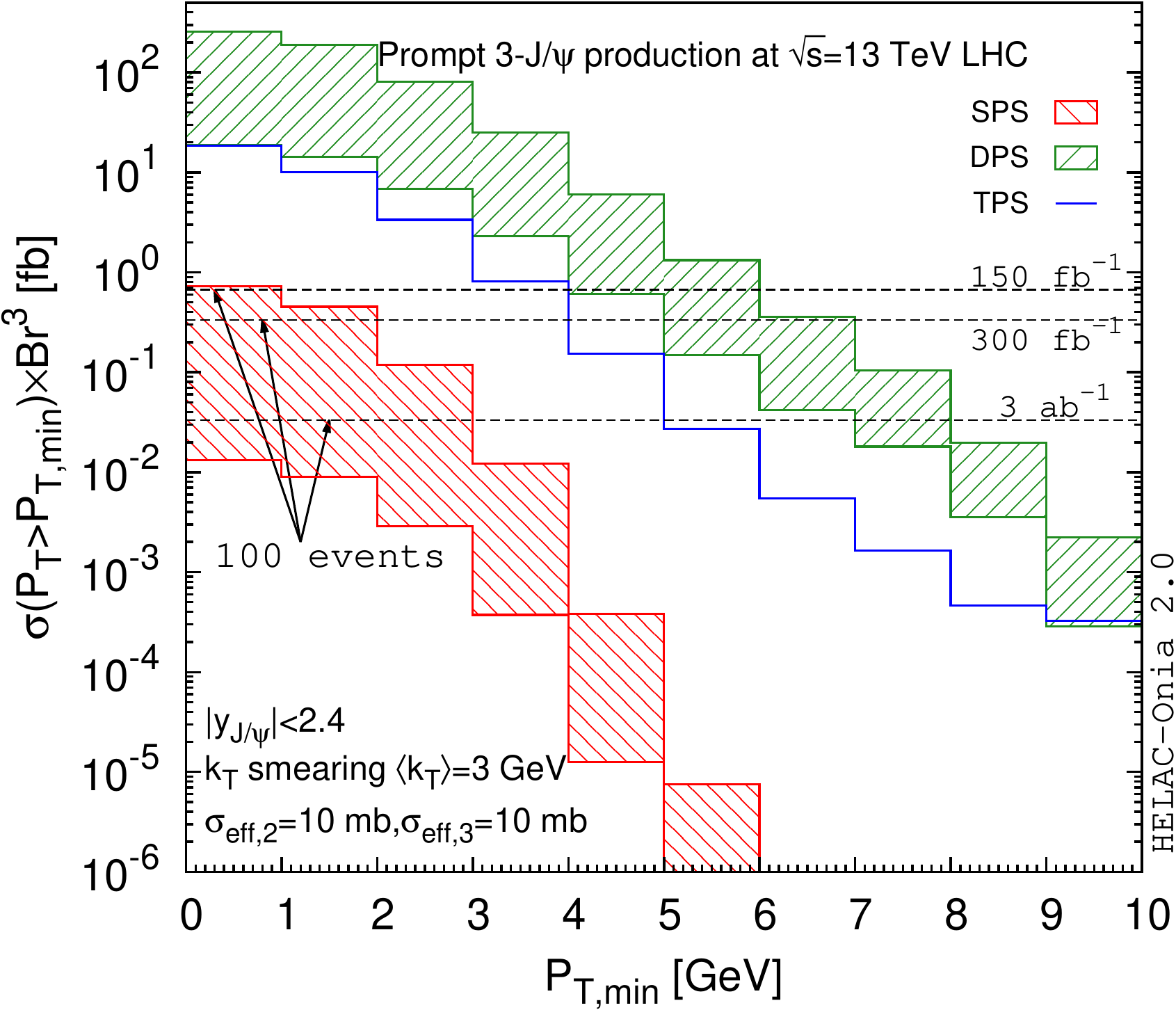}
\caption{Cross section $\sigma(pp\rightarrow 3J/\psi)\times{\rm Br}^3(J/\psi\rightarrow \mu^+\mu^-)$ (in femtobarn) dependence of the minimal transverse momentum cut $P_T>P_{\rm T, min}$ among three $J/\psi$'s at $\sqrt{s}=13$ TeV and within the rapidity interval $|y_{J/\psi}|<2.4$. The horizontal dashed lines are the expected 100 events under targeted integrated luminosities.}
\label{fig:dptCMSplot}
\end{figure}

After fixing $\sigma_{\rm eff,2}=\sigma_{\rm eff,3}=10$ mb, we predict the numbers of triple $J/\psi$ events being $42^{+108}_{-30}$ and $8$ from DPS and TPS, respectively, with the data on tape recorded by the LHCb detector. These numbers will be 50 times higher at the end of the LHC life according to the targeted luminosity. On the other hand, we cannot directly use the numbers in Table~\ref{tab:xsprompt} to estimate the number of events observed by the ATLAS and CMS experiments because of their large magnetic fields and their triggers on the low momentum muons. The lowest transverse momentum $P_T$ of $J/\psi$ that can be observed at these two detectors is not zero. For instance, the minimal $P_T$ of $J/\psi$ in each event is from $4.5$ GeV to $6.5$ GeV in the CMS double $J/\psi$ measurement~\cite{Khachatryan:2014iia}. The cumulative distributions $\sigma(P_T>P_{\rm T,min})\times {\rm Br}^3(J/\psi\rightarrow \mu^+\mu^-)$ are shown in Fig.~\ref{fig:dptCMSplot}, where the SPS, DPS, and TPS cross section are shown individually. $\sigma(P_T>P_{\rm T,min})$ is the cross section with the requirement of $P_T$ of each $J/\psi$ candidate larger than $P_{\rm T,min}$. We have selected events by imposing the rapidity $|y_{J/\psi}|<2.4$ and used a Gaussian distribution with $\langle k_T \rangle=3$ GeV to mimic the (universal) intrinsic $k_T$ smearing effect from the initial states. (The value of $\langle k_T \rangle$ we used here is approximately determined by the LHC measurements of the $J/\psi$ pair~\cite{Aaboud:2016fzt,Aaij:2016bqq}.) Specifically, the whole $k_T$ smearing is assumed to be factorized out by
\begin{eqnarray}
\frac{d\sigma}{d\Phi_{\langle k_T \rangle}}=\int_{0}^{+\infty}{dk_T^2 \frac{\pi}{8\langle k_T\rangle^2}e^{-\frac{\pi}{8}\frac{k_T^2}{\langle k_T \rangle^2}}\frac{d\sigma}{d\Phi}},
\end{eqnarray}
where the phase-space mapping $\Phi  \rightarrow \Phi_{\langle k_T\rangle}$ is determined by boosting the whole event according to the generated transverse momentum imbalance $|\overrightarrow{k}_T|=k_T$ with the uniform azimuthal angle in the transverse plane. The TPS cross section decreases faster than the DPS cross section as $P_{\rm T,min}$ increases. It is understood because TPS has typically a higher-twist effect than DPS. The former is more power suppressed at a higher scale than the latter one. The same argument should be applied to the comparison between SPS and DPS, as the latter is more power suppressed than the former one. It is not the case in Fig.~\ref{fig:dptCMSplot} because of the caveat that we mentioned in the previous section. Like the case of single $J/\psi$ production, the LO calculation in $\alpha_s$ and in $v^2$ is not sufficient to account for the SPS yields at large $P_T$. They might be strongly enhanced by higher-order QCD radiative corrections and the subleading color-octet channels in the same regime. However, given the substantial suppression of SPS compared to DPS and TPS, we do not expect the inclusion of the new channels will significantly change the total yields after summing of the three contributions when $P_T\lesssim 10$ GeV. Three horizontal dashed lines in the figure are indicated for observing 100 events with the integrated luminosities $150$ fb$^{-1}$ (data on tape), 300 fb$^{-1}$ (phase 1 of HL-LHC) and 3 ab$^{-1}$ (phase 2 of HL-LHC), respectively. With $150$ fb$^{-1}$, one is able to observe more than 100 selected events with $P_T>5$ GeV. The Monte Carlo simulations are necessary in order to properly take into account the realistic experimental conditions, like the trigger, reconstruction efficiency, and combinatorial backgrounds. Such a study is beyond the scope of our Letter.

\begin{figure}[hbt!]
\centering
\includegraphics[width=.95\columnwidth,draft=false]{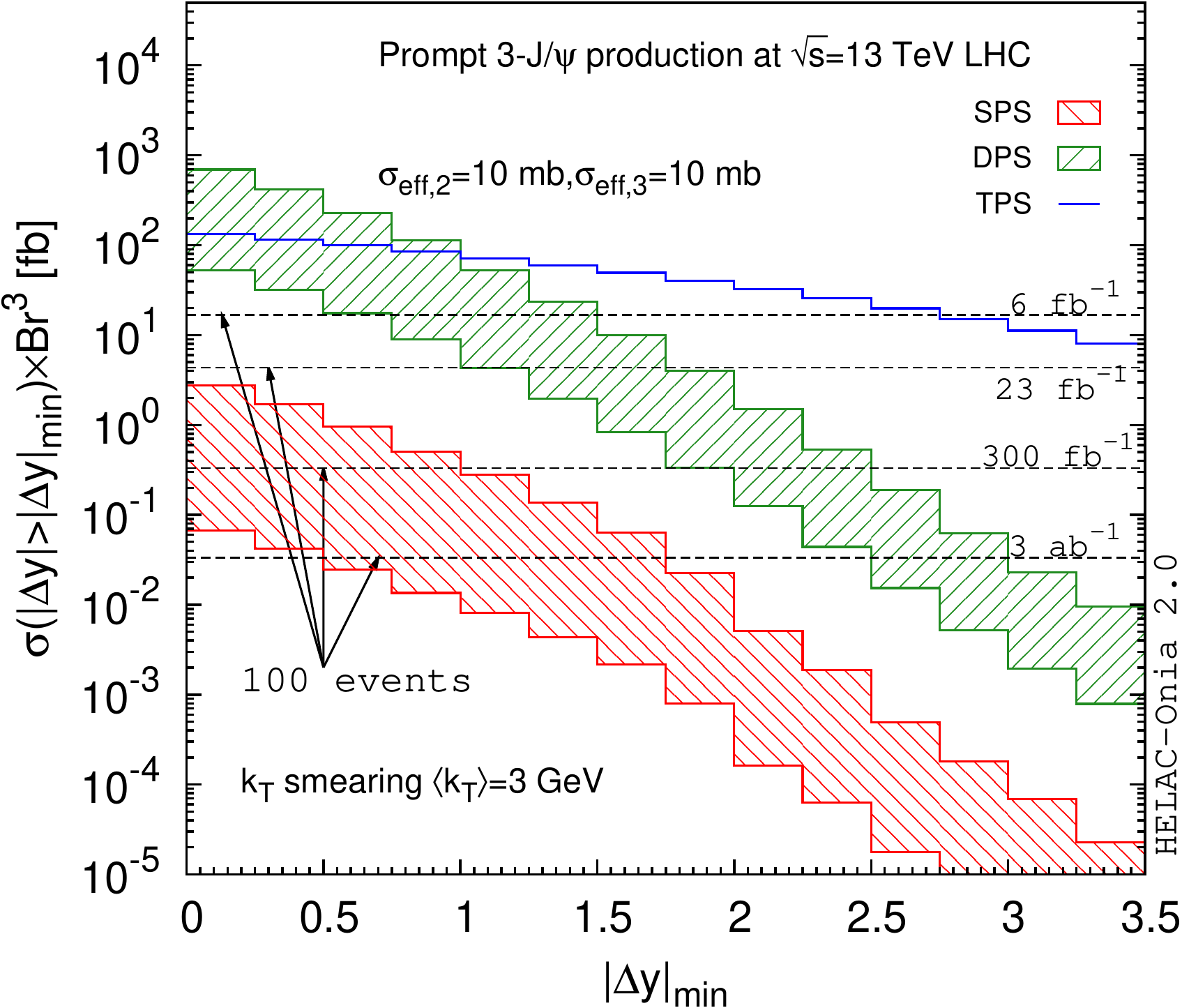}
\caption{Cross section $\sigma(pp\rightarrow 3J/\psi)\times{\rm Br}^3(J/\psi\rightarrow \mu^+\mu^-)$ (in femtobarn) dependence of the minimal rapidity gap cut $|\Delta y|>|\Delta y|_{\rm min}$ among three $J/\psi$'s at $\sqrt{s}=13$ TeV. The horizontal dashed lines are the expected 100 events under targeted integrated luminosities.}
\label{fig:dyplot}
\end{figure}

In order to filter out the TPS events, a good observable is to use the minimal rapidity gap among the three $J/\psi$ mesons. Such an observable has the virtue of being insensitive to the intrinsic $k_T$ smearing, as opposed to other observables like the azimuthal angles. The cumulative distributions $\sigma(|\Delta y|>|\Delta y|_{\rm min})\times {\rm Br}^3(J/\psi\rightarrow \mu^+\mu^-)$ can be found in Fig.~\ref{fig:dyplot}, where $|\Delta y|$ is the minimal absolute rapidity difference among the three possible combinations of a $J/\psi$ pair. Because none of the three $J/\psi$ pairs are correlated in TPS, in contrast to SPS and DPS, it has higher possibility of generating an event with a large rapidity gap. Indeed, TPS contribution starts to be dominant when $|\Delta y|>1$. The situation here is quite similar to the absolute rapidity difference in the double $J/\psi$ production, which has been extensively used to extract DPS in the process. The minimal rapidity gap $|\Delta y|$ can be readily used to determine TPS information in the triple $J/\psi$ production. Thus, it paves the way to study the triple-parton correlations in a proton for the first time via the TPS triple $J/\psi$ process.

\medskip


\textit{Conclusions} -- We have proposed to use triple prompt $J/\psi$ production at the LHC and the future hadron colliders to improve our knowledge of the multiple-parton scattering physics. In particular, the TPS has never been observed in experiments. The triple prompt $J/\psi$ hadroproduction can be a very clean process to probe TPS and, therefore, the possible triple-parton correlations in a proton. We performed a first complete theoretical study of the process by including SPS, DPS, and TPS contributions. Especially, we have accomplished the very challenging task of the PQCD calculation for triple $J/\psi$ SPS production at $\mathcal{O}(\alpha_s^7)$, which involves more than $2\cdot 10^{4}$ Feynman diagrams. Our calculation shows that it is a DPS and TPS dominant process, and therefore a golden channel to probe MPI. Although the process is rare, we have shown that the data on tape after LHC run 2 is already more than enough to measure the process. Finally, we also pointed out that the minimal rapidity gap among three $J/\psi$'s is a very useful observable to separate the TPS events from the DPS and SPS events.

\acknowledgments
H.-S.S is supported by
the ILP Labex (ANR-11-IDEX-0004-02, ANR-10-LABX-63). Y.-J.Z is supported by the National Natural Science Foundation of China (Grants No. 11722539).

\bibliography{paper}

%
%
%
%
%

\end{document}